\documentstyle[preprint,aps]{revtex}
\begin{document}
\draft
\baselineskip 24 true pt
\vsize=9.5 true in \voffset=.2 true in
\hsize=6.0 true in \hoffset=.25 true in

\newpage
\begin{center}
{\Large{\bf Magnetic transition and polaron crossover in a two-site single polaron model including double exchange interaction}}
\end{center}
\vskip 1.0cm
\begin{center}
\end{center}
\vskip 0.50cm
\begin{center}
{Jayita Chatterjee\footnote{ e-mail: moon@cmp.saha.ernet.in}, Manidipa Mitra 
and A. N. Das} 
\end{center}

\begin{center}
 {\em Saha Institute of Nuclear Physics \\
1/AF Bidhannagar, Calcutta 700064, India}\\

\end{center}

\vskip 1.0cm

\vskip 1.0cm
\begin{center}
{\bf Abstract}
\end{center}
\vskip 0.5cm
A two-site double exchange model with a single polaron is studied 
using a perturbation expansion based on the modified Lang-Firsov 
transformation. The antiferromagnetic to ferromagnetic transition 
and the crossover from small to large polaron are investigated 
for different values of the antiferromagnetic 
interaction ($J$) between the core spins and the hopping ($t$) of the 
itinerant electron. Effect of the external magnetic field on the 
small to large polaron crossover and on the polaronic kinetic energy  
are studied. When the magnetic transition and 
the small to large polaron crossover coincide for some suitable 
range of $J/t$, the magnetic field has very pronounced effect 
on the dynamics of polarons.

\newpage
\begin{center}
{\bf 1. Introduction}   
\end{center}
\vskip 0.3cm 
The origin of ferromagnetism in manganese perovskites 
$La_{1-x}X_xMnO_3$ ($X = Ba$, $Sr$, $Ca$ etc.) is the double exchange
mechanism \cite{zen,gen}. The discovery of anomalous magnetotransport 
phenomena \cite{von} in these
compounds has stimulated intensive studies in its magnetic as well as
electrical properties. However, the simple double-exchange interaction
alone is not sufficient to explain the experimental results \cite{von}.
Depending on the doping, temperature and the radius of the dopant ion
these oxides show various phases \cite{rao}. A complete understanding
of the properties of $Mn$ perovskites is still far from clear.

One of the keys to understand unusual physical properties is to find the
role of coupling between the carriers and the underlying lattice. Several
theoretical models have been proposed based on lattice-carrier
coupling \cite {rod,lee,yu2,Dg,Yu,yar}. Many recent experiments
\cite{zhao} indicate that the electron-phonon ($e$-ph) interaction shapes
its properties very crucially. Moreover, small to large polaron
crossover is reported by many experimental groups \cite{bill,louc,booth,lanz}.
There are models \cite{lee,yu2,yar} which incorporate double exchange
interaction in a polaronic model. Min and co-workers \cite{lee}
have studied the role of $e$-ph interaction in systems where double
exchange interaction is present. The combined model of spin double exchange
and lattice polaron \cite{lee} is used to investigate the effect
of small to large polaron crossover on the magnetic and transport properties
under the mean-field approximation scheme. It concludes that the effect of 
polaron narrowing on the colossal magnetoresistance (CMR) is
more pronounced in CMR manganites with
low magnetic transition temperature ($T_c$) than in the 
high $T_c$ manganites. Within the mean field theory \cite{lee} 
the magnetic transition, metal-insulator transition
 and a large drop in magnetoresistance occur at the same temperature.

In an attempt to have a clearer view from a nearly exact calculation,
we include double exchange interaction in a two site one polaron model 
and follow a perturbation expansion \cite{jayee} based on a modified 
Lang-Firsov (MLF) phonon basis where the lattice distortions 
produced by the electron are treated as variational 
parameters \cite {DS,DC,LS}. 
This method shows good convergence as well as nearly exact results \cite{jayee} 
for almost the entire range of $e$-ph coupling 
for $t/\omega_0 \le 1$, where $\omega_0$ is the phonon frequency. 
We investigate the ferromagnetic (FM) to antiferromagnetic (AFM) 
transition, the crossover from large to small polaron, the 
behavior of the effective hopping of the itinerant electron and 
the kinetic energy as a 
function of the $e$-ph coupling strength for the ground state 
of the system. 
Effect of the magnetic field on the large to small polaron 
crossover and on the polaronic kinetic energy are also studied. 

The paper is organized as follows. In Sec. 2 we define the model 
hamiltonian describing different interactions  
and calculate different physical quantities which indicate the behaviour of 
small-large polaron crossover and AFM-FM transition as a function of 
$e$-ph coupling. In Sec. 3 we present the results obtained in
our calculation and discussions. Sec. 4 contains the concluding remarks.   
\vskip 1.0cm

\begin{center}
{\bf 2. Formalism }
\end{center} 
\vskip 0.3cm

The Holstein model \cite{Hol} including double exchange interaction for the two-site single-polaron system is described by the Hamiltonian
\begin{eqnarray}
H &=& \sum_{i,\sigma} \epsilon n_{i \sigma} 
- \sum_{\sigma} t \cos({\frac{\theta}{2}})(c_{1\sigma}^{\dag} c_{2 \sigma}+ 
c_{2 \sigma}^{\dag} c_{1 \sigma})\nonumber\\
&+& g \omega_0  \sum_{i,\sigma}  n_{i \sigma} (b_i + b_i^{\dag}) 
+  \omega_0 \sum_{i}  b_i^{\dag} b_i + J\sum_{<ij>} \vec S_{i}.\vec S_{j} 
\end{eqnarray} 
where $i$ =1 or 2, denotes the site. $c_{i\sigma}$ ($c_{i\sigma}^{\dag}$)  
is the annihilation (creation) operator for the itinerant electron with
spin 
$\sigma$ at site $i$, $n_{i \sigma}$ (=$c_{i\sigma}^{\dag} c_{i\sigma}$) 
is the corresponding number operator and $g$ denotes the on-site $e$-ph coupling
strength. $b_i$ and $b_{i}^{\dag}$ are the annihilation and 
creation operators respectively for the phonons corresponding to 
interatomic vibrations at site $i$, $\omega_0$ is the corresponding phonon 
frequency and $\epsilon$ is the bare-site energy. 
$\vec S_{i}$ and $\vec S_{j}$ are the core-spins at the site $i$ and $j$
respectively, $\theta$ is the angle between the core-spins $\vec S_{i}$ and 
$\vec S_{j}$.  
The transfer hopping integral ($t$) is modified through the relative 
angle $\theta$ as 
$t\cos({\frac{\theta}{2}})$ because of the strong Hund's coupling 
between the spins of the core eletrons and itinerant
electron \cite{gen}. 
$J$ is the superexchange antiferromagnetic interaction 
between the neighbouring core-spins.  
Since we will restrict ourselves to the single-electron case we will 
not consider the electron spin indices. 

If one compares our simple model 
with the manganite (e.g. $La_{1-x}Ca_{x}MnO_3$) system the core spin and the
itinerant electron may be identified with the $t_{2g}^3$ (localized) electrons 
and the $e_{g}^1$ (mobile) electron of $Mn^{3+}$ of the manganite system 
respectively. 
It may be mentioned that for manganites 
the Jahn-Teller(JT) coupling is important whereas 
the Holstein phonon mode, in general, 
should correspond to the breathing mode in manganites \cite{Dg}.  
To include JT coupling one should consider two $e_g$ orbitals 
(say $\alpha$ and $\beta$) and their coupling to the JT phonon mode. For 
the single electron case, as considered here, if one neglects the 
interorbital hopping ($t_{\alpha\beta}=0$) then 
the Hamiltonian can be separated into two parts each corresponds to 1-orbital
case and as a consequence the 2-orbital problem 
is reduced to an effective single orbital problem 
with an effective Holstein type interaction as in Hamiltonian (1).
For the two electron case the above simplification cannot be done and
a study of two-site two-electron problem considering two orbitals in 
the context of manganites is in progress.
We have considered here one itinerant electron in a two-site 1-orbital system.
So, the density of itinerant electron $<n>$ is equal to 0.5 which, in principle,corresponds to $x=0.5$ for manganite system. 
It may be mentioned that the studies of 1-orbital \cite {Yu} and 
2-orbital Kondo model \cite{yu2} in the context of manganites  
using Monte Carlo techniques showed that the results for 
2-orbital model at $n=1$ ($x=0$) are similar to 
those for single orbital model at $x=0.5$. 

By introducing new 
phonon operators $a=~(b_1+b_2)/ \sqrt 2$ and $d=~(b_1-b_2)/\sqrt 2 $, 
the  Hamiltonian ($H$) is separated into two parts :
one corresponding to the in-phase mode which does not couple with the
electronic degrees of freedom and the other involving out-of-phase mode 
($H_d$) which represents an effective $e$-ph system and cannot be 
solved analytically \cite{RT}.  

The MLF transformation with variable phonon basis is used  
so that a convergent perturbation expansion can be obtained. 
The transformed Hamiltonian is 
\begin{eqnarray}
\tilde{H_d}&=& e^R H_d e^{-R}\nonumber\\
&=&\omega_0  d^{\dag} d + \sum_{i} \epsilon_p n_{i} - 
t\cos({\frac{\theta}{2}})~ [c_{1}^{\dag} c_{2}~ \exp
(2 \lambda (d^{\dag}-d)) \nonumber\\    
&+& c_{2}^{\dag} c_{1}~\exp(-2 \lambda (d^{\dag}-d))]  
+ \omega_0 (g_+ - \lambda)(n_1 - n_2)(d + d^{\dag}) 
+JS^2 \cos{\theta}
\end{eqnarray}
where $R =\lambda (n_1-n_2) ( d^{\dag}-d)$, $\lambda$ is a
variational parameter related to the displacement of the $d$ 
oscillator, $g_{+}=g/\sqrt 2$
and $\epsilon_p = \epsilon - \omega_0 ( 2 g_{+} - \lambda) \lambda$. 
              
For the perturbative expansion following Ref. \cite{jayee} the basis set 
is chosen as 
$|\pm,N \rangle = \frac{1}{\sqrt 2} (c_{1}^{\dag} \pm  c_{2}^{\dag})$
$|0\rangle_e  |N\rangle$,    
where $|+\rangle$ and $|-\rangle$ are the bonding and antibonding 
electronic states and $|N\rangle$ denotes the $N$th excited oscillator  
state within the MLF phonon basis. 
The diagonal part of the Hamiltonian $\tilde{H_d}$ in the chosen basis 
is treated as the unperturbed Hamiltonian ($H_0$) 
and the 
remaining part of the Hamiltonian $H_{1}= \tilde{H_{d}}-H_0$,  
as a perturbation. 

The unperturbed energy of the state $| \pm,N\rangle$ is given by 
\begin{eqnarray}
 E_{\pm,N}^{(0)}&=& \langle N,\pm|H_0|\pm, N \rangle\nonumber\\
&=& N \omega_0  + \epsilon_p \mp t_{eff} \left[ \sum_{i=0}^{N}
 \frac{(2\lambda)^{2i}}{i!} (-1)^i N_{C_i}\right] +JS^2 \cos{\theta}
\end{eqnarray}
where $t_{eff}=t~ \cos{\frac{\theta}{2}}~\exp{(-2\lambda^2)}$ and 
the general off-diagonal matrix elements of 
$H_1$ between the two states $|\pm,N \rangle$ and $|\pm,M \rangle$ 
may be calculated for $(N-M)>0$ as in Ref. \cite{jayee}.

Within the chosen basis, the unperturbed ground state is the 
$|+\rangle|0\rangle$ state 
with the unperturbed energy, $ E_0^{(0)}=\epsilon_p-t_{eff}+  
JS^2 \cos{\theta}$.  
 
The first order correction to the ground state wave function is 
obtained as,  
\begin{eqnarray}  
|\psi_0^{(1)}\rangle =\frac{[\omega_0(g_{+}-\lambda)-
2\lambda t_{eff}]}{(E_{0}^{(0)}-E_{+,1}^{(0)})}~|-,1\rangle 
- \sum_{N=2,3,4,..}\frac{t_{eff}(2\lambda)^N}{\sqrt{N!}
( E_0^{(0)}-E_{e,N}^{(0)})}~|e,N\rangle 
\end{eqnarray}  
where e = + or - for even and odd N respectively.

The first order correction to the energy ($E_0^{(1)}$) is zero since 
$H_{1}$ has no diagonal matrix element in the chosen basis. 
The second order correction to the ground state energy 
is given by   
\begin{eqnarray}
E_0^{(2)} &=& \sum_{N=1,3..}\frac{|\frac{-t_{eff}(2\lambda)^{N}}
{\sqrt{N!}}+\omega_0 (g_+ - \lambda)~\delta_{N,1}|^2}
{(E_0^{(0)}-E_{-,N}^{(0)})}\nonumber\\ 
&+&\sum_{N=2,4..}\frac{|\frac{-t_{eff}(2\lambda)^{N}}
{\sqrt{N!}} |^2}{(E_0^{(0)}-E_{+,N}^{(0)})} 
\end{eqnarray}
Higher order corrections to the ground state wavefunction and energy
are obtained following Ref. \cite{jayee}. For the study of the effect 
of the magnetic field ($h$) we include a term 
-$\tilde{g}\sum_{i} \mu_B hS_{i} \cos {\frac{\theta}{2}}$ to the Hamiltonian in Eq. (1) (where $\tilde{g}$ is the Lande g factor) and as a
result a term $-2 \mu_{eff} h \cos{\frac{\theta}{2}}$ is added to 
$E_0^{(0)}$ where $\mu_{eff}$ (= $\tilde{g}S\mu_{B}$) is the local 
moment of the core spins. In this paper we express the magnetic field
($h$) in a unit of $\mu_{eff}=1$.  
 
 Now a proper choice of $\lambda$ is to be made so that the 
perturbative expansion becomes convergent. 
Our previous work \cite{jayee} has shown that the $\lambda$, 
obtained by minimizing the unperturbed ground state energy, gives
satisfactory convergence to the perturbation series for $t/\omega_0 \le 1$ 
and the convergence becomes very rapid and excellent with 
decreasing value of $t/\omega_0$. Here also we will follow the 
similar procedure. Minimizing 
the unperturbed ground state energy $E_0^{(0)}$ with respect 
to $\lambda$ we obtain 
\begin{eqnarray}
\lambda&=&\frac {\omega_0g_{+}}{\omega_0+2t_{eff} }  
\end{eqnarray}
Minimization of the unperturbed ground state energy with respect to 
$\theta$ gives an approximate value ($\theta_{\rm MLF}$) of $\theta$
\begin{eqnarray}
\cos{\frac{\theta_{\rm MLF}}{2}}& = & [\frac{t \exp{(-2\lambda^2)}}{4JS^2}
+\frac{\mu_{eff} h}{2JS^2}] ~~\rm{(for~ nonzero ~solution ~of~} 
\theta_{\rm MLF})\nonumber 
\end{eqnarray}
However, exact $\theta$ should be evaluated from the minimization of
the exact ground state energy. For each value of $g_+$ we calculate 
the energy upto the sixth order in perturbation \cite{jayee} and find 
out for which value of $\theta$ the energy (including perturbation 
corrections) is minimum.  

 We have also calculated $t_{eff}^{KE}=-E_{Kin}=<\psi_G|$
$t\cos({\frac{\theta}{2}})~ [c_{1}^{\dag} c_{2}~ \exp$
$(2 \lambda (d^{\dag}-d))$ 
$+ c_{2}^{\dag} c_{1}~\exp(-2 \lambda (d^{\dag}-d))]$
$|\psi_G>$ where $\psi_{G}$ is the ground state wave-funcion which we have 
calculated upto the fifth order corrections in perturbation. $t_{eff}^{KE}$
describes the kinetic energy of the system and it reduces to the effective 
hopping ($t_{eff}$) for both the weak and strong
coupling limits \cite{mello}.

   The static correlation functions $\langle n_1 u_{1}\rangle_{0}$ and 
$\langle n_1 u_{2}\rangle_{0}$,
where $u_1$ and $u_2$ are the lattice deformations at sites 1 and 2  
respectively, produced by an electron at site 1, are the standard 
measure of polaronic character. 
The correlation functions for this two-site system may be 
written as \cite{jayee}
\begin{eqnarray}
\langle n_{1} u_{1} \rangle_{0}&=&\frac{1}{2}  
\left[-(g_{+} +\lambda) + \frac{A_0}{N_G}\right] \\
\langle n_{1} u_{2} \rangle_{0}&=&\frac{1}{2}
\left[-(g_{+}-\lambda)- \frac{A_0}{N_G}\right] \nonumber
\end{eqnarray}  
\begin{eqnarray} 
{\rm where}~~~A_0\equiv\langle \psi_{G} |n_1(d+d^{\dag})|\psi_{G}
\rangle \nonumber
\end{eqnarray} 
where $N_{G}$ is the normalization factor 
to the wavefunction. The variation of the physical quantity 
$\lambda_{corr}/g_{+}~ =~- <n_1(u_1-u_2)>_0/g_+$ with the 
$e$-ph coupling strength manifests 
the nature of large to small polaron crossover. 
Its value becomes 1 in the extreme small polaronic limit where the 
lattice distortion is very local and decreases
significantly in the small to large polaron crossover region as $g_+$ 
decreases.
\vskip 3.5cm
\begin{center}
{\bf 3. Results and Discussions}
\end{center}
\vskip 0.3cm
 
 In the combined model of polaron and double exchange interaction, the 
nature of variation of $\theta$, $t_{eff}^{KE}$ and $\lambda_{corr}/g_+$ 
with $g_+$ 
reveal different transitions and crossover regions. We focus our 
attention to the parameter space where $t>J$ as it corresponds to 
real systems \cite{yu2,Per}. We present the results for $t=1.0$ and 
$JS^2=0.05, 0.15$, 0.25 and $0.5$ (in a scale of $\omega_0=1.0$).  
The bare-site energy $\epsilon$ is taken as zero. 

Fig. 1 shows the change of relative orientation ($\theta$) of two 
core spins as a function of $e$-ph coupling strength ($g_+$) for the
ground state. For small values of $g_+$ two core 
spins are either aligned parallel or canted depending on the strength 
of $t/JS^2$. With increasing 
$g_+$ the angle ($\theta$) increases and finally a transition from the 
FM or canted AFM state to the AFM
state occurs. Nature of this
transition depends on the ratio of $t$ to $JS^2$. For $t$=1.0, 
$JS^2$=0.05 the FM to AFM transition is very sharp. With increasing value 
of $JS^2$ the transition occurs at a lower value of $g_+$ and it becomes 
broader. 
For higher values of $JS^2$ (=0.25, 0.5), the crossover from the 
canted state to the AFM state is very smooth. It may be mentioned that 
Yunoki and Moreo \cite{Yu} studied the 1-orbital Kondo model with 
AF interaction ($J$) between the core spins (in absence of $e$-ph interaction) 
in the context of manganites and obtained FM state at $n=0.5$ for low 
values of $J$. With increasing $J$ the ground state at quarter 
filling ($n=0.5$) appears to be a spin 
spiral state and ultimately an AF state. For the two orbital model in presence 
of $e$-ph coupling a similar FM state is observed for low values of 
$e$-ph coupling for $n=1$ and an AF state for large values of $e$-ph coupling.

 With increasing $g_+$ a large to small 
polaron crossover occurs in the system in addition to the magnetic transition. 
This is clearly seen from Figs. 2 and 3 where the variations of $t_{eff}^{KE}$, $\lambda_{corr}/g_+$ and $\theta$ with $g_+$ are shown.  
The large to small polaron crossover is identified by the sharp
fall in $t_{eff}^{KE}$ and rise in $\lambda_{corr}$. 
For $t=1.0 $ and $JS^2=0.15$ the FM to AFM transition and the 
large to small polaron crossover occur simultaneously (Fig. 2) around 
$g_+ =1.2$. However, for $JS^2$=0.05 the magnetic transition 
occurs at a higher values of $g_+$ than that coresponding to 
large to small polaron crossover (Fig. 3). 
In this situation the FM state has two regions: one with appreciable 
value of $t_{eff}^{KE}$ and the other with much reduced value 
of $t_{eff}^{KE}$ (Fig. 3). In the former case, charge carrier is the 
delocalized large polaron with appreciable hopping and this regime 
is expected to be metallic in the thermodynamic limit. In the latter 
region the hopping would be heavily 
suppressed by the small polaron formation and it may 
result in an insulating FM state in some cases. 
Thus we find that there is a 
possibility of FM metal-FM insulator-AFM insulator transition 
for low values of $JS^2/t$.  
It is to be noted that if one follows the usual MLF method 
(zeroth order of perturbation within our approach) the FM-AFM 
transition coincides with large to small polaron crossover
for any value of $JS^2/t$. This is demonstrated in Fig. 3 for 
$JS^2/t$=0.05, where we find the coincidence of the 
FM-AFM transion with the large-small polaron crossover within the usual
MLF approach whereas the nearly exact calculations yield different
result. The nearly exact results, obtained from the MLF perturbation  
method, makes the large-small polaron crossover smooth
in favour of the conclusion of L\"owen \cite{L}.

 In Fig. 4 we have shown the effect of the external magnetic 
field on $\theta$, $\lambda_{corr}/g_+$ and $t_{eff}^{KE}$ for 
$JS^2=0.15$.  
The magnetic 
field favours the FM state, hence the FM-AFM transition and associated 
large-small polaron crossover (for $JS^2/t=0.15$) take place at a higher
value
of $g_+$ with increasing field. From Fig. 4 it is clear that 
the magnetic field reduces the value of
the local distortion ($\lambda_{corr}/g_+$) and enhances $t_{eff}^{KE}$ 
significantly in the large-small polaron crossover
region provided the crossover is associated with a FM-AFM transition. 

For $JS^2/t=0.05$ the system remains in the FM state in the large-small polaron
crossover region, consequently the magnetic field 
has no effect on $t_{eff}^{KE}$ in this crossover region. However, 
the field has an effect on $\theta$, hence on $t_{eff}^{KE}$ at the FM-AFM
transition.
  
In Fig. 5 we plot the change in $t_{eff}^{KE}$ due to the magnetic field
as a function of $g_+$. This quantity may be related to the 
magnetoresistance for a system in the thermodynamic limit. 
For polarons the conductiviy at low temperatures is dominated by 
the tunneling mobility which is proportional to $t_{eff}^2$ within 
the zeroth order of perturbation \cite{yar}. In general the $t_{eff}^{KE}$ 
is a measure of delocalization of the electron. A reduction 
of $t_{eff}^{KE}$ will cause a reduction in mobility.  
In Fig. 5 the change in $t_{eff}^{KE}$ due to the field is very sharp and 
prominent at the FM-AFM transtion when it is associated with 
the large-small polaron crossover for $JS^2/t=0.15$, 
while it shows a broad smaller peak for $JS^2/t=0.05$ where the 
FM-AFM transition is not associated with the polaron crossover. 
For the former case the magnetic field would have pronounced
effect on the transport properties.

\vskip 3.0cm
\begin{center}
{\bf 4. Conclusions}
\end{center}
\vskip 0.3cm

 From our studies on the two-site double exchange model with a single
polaron as a function of $e$-ph coupling strength we conclude that the 
nature of the FM-AFM transition depends on the relative values of 
$J$ and $t$. The transition is sharper for smaller values of $J/t$. 
For high values of $J/t$   
a canted state is stable instead of a FM state for weak $e$-ph coupling ($g_+$) 
and the crossover from the 
canted state to the AFM state is very smooth with increasing $g_+$.   
For suitable values of $J/t$ (=0.15) the FM-AFM transition coincides 
with the large-small polaron crossover. For this case the external
magnetic field has very prominent effect on the polaronic local 
distortion and kinetic energy in the transition region 
(as in Figs. 4(b) and 4(c) respectively). 
Within the MLF method (zeroth order of perturbation) the polaron
crossover is always associated with FM-AFM transition for any value 
of $J/t$ whereas our results based on convergent perturbation expansion
yield that the coincidence of magnetic transition and polaron crossover
will depend on the value of $t$ and $J$.  
For low values of $J/t$ a crossover from the FM large 
polaronic state to the FM small 
polaronic state with reduced hopping occurs and then 
a transition to the AFM state takes place at a higher value 
of $g_+$.  


\newpage
Figure captions :
\noindent

FIG. 1. Variation of the relative angle ($\theta$) between the core-spins,  
with $g_+$ for $t=1.0 $ and $JS^2=0.05$, 0.15, 0.25 and 0.5 (in units of
$\omega_0$=1). 
\vskip 0.5 cm
\noindent

FIG. 2. Variations of $\theta$, $t_{eff}^{KE}$ and 
$\lambda_{corr}/g_+$ with $g_+$ for $t=1.0$, $JS^2=0.15$ and $h=0$. 
\vskip 0.5cm
\noindent

FIG. 3. Variations of $\theta$ (with and without perturbation), $t_{eff}^{KE}$, $t_{eff}$, $\lambda_{corr}/g_+$ and $\lambda/g_{+}$ 
with $g_+$ for $t=1.0$, $JS^2=0.05$ and $h=0$.
Note that $\theta$(MLF), $t_{eff}$ and $\lambda/g_+$ are obtained
within the MLF method without considering perturbation corrections.
\vskip 0.5cm

FIG. 4. Variation of (a) $\theta$, (b) $\lambda_{corr}/g_+$ and (c)
$t_{eff}^{KE}$ with $g_+$ for different values of the magnetic field 
$h=$0, 0.02 and 0.04 (in units of $\mu_{eff}$=1) for $t=1.0 $ and 
$JS^2=0.15$. 

\vskip 0.5 cm
\noindent

FIG. 5. Variation of the change in $t_{eff}^{KE}$ due to the magnetic field 
($\frac{(t_{eff}^{KE}(h)- t_{eff}^{KE}(0))}{h}$) 
as a function of $g_+$ 
 for $t=1.0 $, $JS^2=0.05$ and 0.15 and $h$=0.04.
\vskip 0.5 cm
\noindent


\begin{thebibliography}{999}

\bibitem{zen}{C. Zener, Phys. Rev. {\bf 82}, 403 (1951)} 

\bibitem{gen}{P. W. Anderson and H. Hasegawa, Phys. Rev, {\bf 100}, 675 (1955); P. G. de Gennes, Phys. Rev. {\bf 118}, 141 (1960).} 

\bibitem{von} {R. von Helmolt, J. Wecker, B. Holzapfel, L. Schultz and
K. Samwer, Phys. Rev. Lett. {\bf 71}, 2331 (1993). }

\bibitem{rao}{A. P. Ramirez, J. Phys. : Condensed matter {\bf 9}, 8171 (1997);
C. N. R. Rao, A. Arulraj, A. K. Cheetham and B. Raveau,
J. Phys : Condensed matter {\bf 12}, R83 (2000)}

\bibitem{rod}{H. R\"{o}der et al, Phys. Rev. Lett. {\bf 76}, 1356 (1996); 
A. J. Millis et al {\it ibid} {\bf 77}, 175 (1996).}

\bibitem{lee}{J. D. Lee and B. I. Min, Phys. Rev B {\bf 55}, 12454 (1997); 
Unjong Yu and B. I. Min cond-mat/9906263.}

\bibitem{yu2}{S. Yunoki, A. Moreo and E. Dagotto, Phys. Rev. Lett. {\bf 81}, 5612 (1998).}

\bibitem{Dg}{T. Hotta, S. Yunoki, M. Mayr and E. Dagotto, 
Phys. Rev. B {\bf 60}, R15009 (1999).}

\bibitem{Yu}{S. Yunoki and A. Moreo, Phys. Rev. B {\bf 58}, 6403 (1998).}

\bibitem{yar}{A. S. Alexandrov and A. M. Bratkovsky, J. Phys.: 
Condensed matter {\bf 11}, 1989 (1999)}

\bibitem{zhao}{G. Zhao et al, Nature {\bf 381}, 676 (1996).}

\bibitem{bill}{S. J. L. Billinge et al Phys. Rev. Lett. {\bf 77}, 715 (1996).}

\bibitem{louc}{D. Louca et al Phys. Rev. B {\bf 56}, R8475 (1997).}

\bibitem{booth}{C. H. Booth et al, Phys. Rev. Lett. {\bf 80}, 853 (1998).}

\bibitem{lanz}{ A. Lanzara et al, Phys. Rev. Lett. {\bf 81}, 878 (1998).}

\bibitem{jayee}{A. N. Das and Jayita Chatterjee, Int. Jour. Mod. Phys.,
{\bf 13}, 3903 (1999); Jayita Chaterjee and A. N. Das, Phys. Rev. B
{\bf 61}, 4592 (2000).}

\bibitem{DS}{A. N. Das and S. Sil, Phyica C {\bf 207}, 51 (1993); 
J. Phys.: Condens. Matter {\bf 5}, 1 (1993).}

\bibitem{DC}{A. N. Das and P. Choudhury, Phys. Rev. B {\bf 49}, 
13219 (1994).}

\bibitem{LS}{C. F. Lo and R. Sollie, Phys. Rev. B {\bf 45}, 7102 (1992).} 

\bibitem{Hol}{T. Holstein, Ann. Phys. (NY) {\bf 8}, 325 (1959).}

\bibitem{RT}{J. Ranninger and U. Thibblin, Phys. Rev. B {\bf 45}, 7730 
(1992).}

\bibitem{mello}{E. V. L. de Mello and J. Ranninger, Phys. Rev. B {\bf 55}, 
14872 (1997).}

\bibitem{Per}{T. G. Perrin et al, Phys. Rev. Lett. {\bf 78}, 3197 (1997).} 

\bibitem{L}{H. L\"owen, Phys. Rev. B {\bf 37}, 8661 (1988).}
\end{thebibliography}
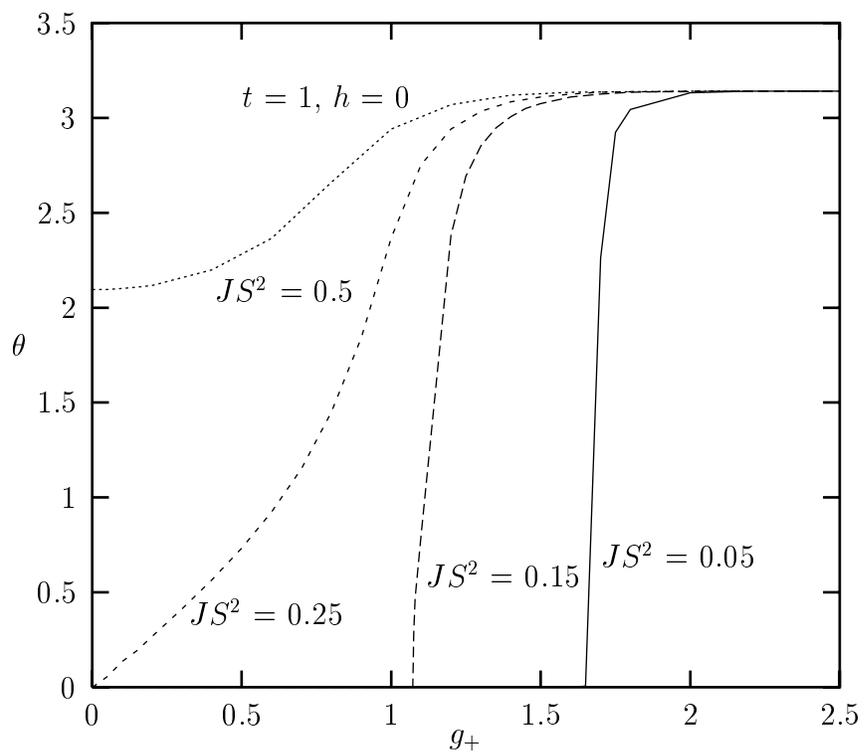
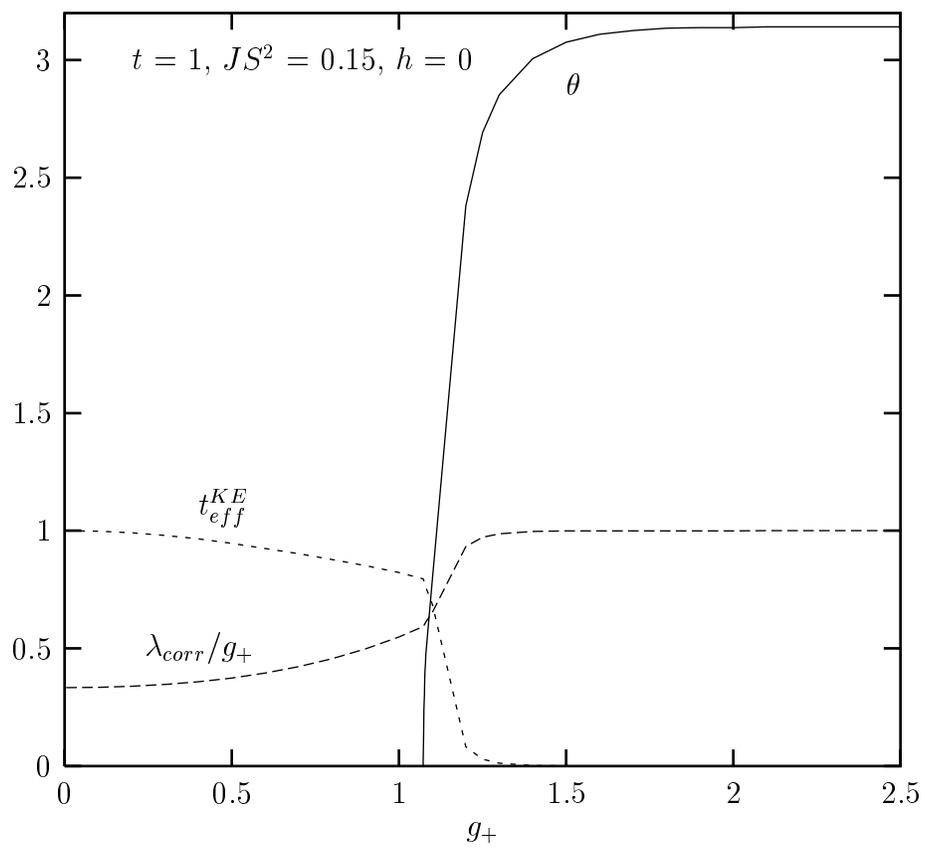
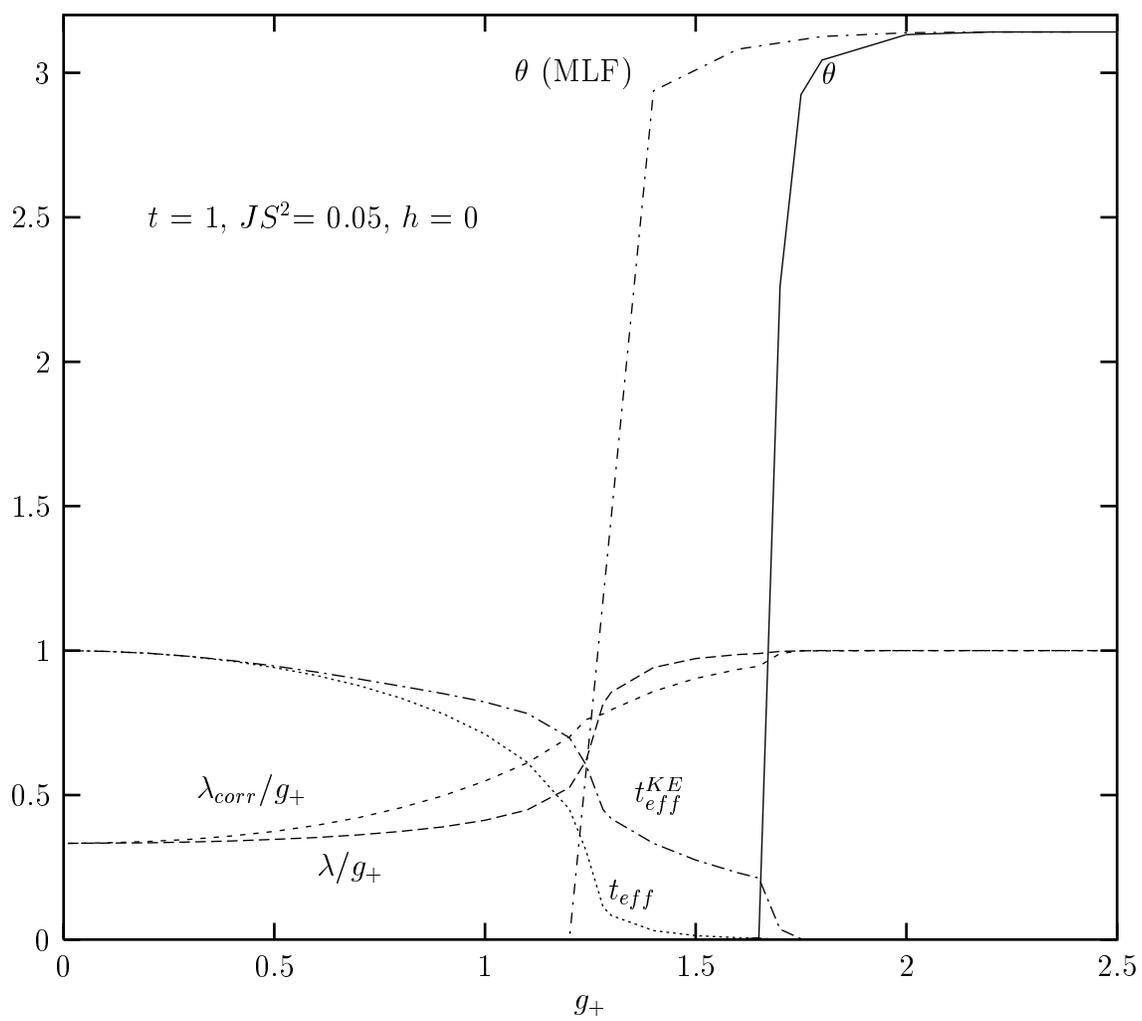
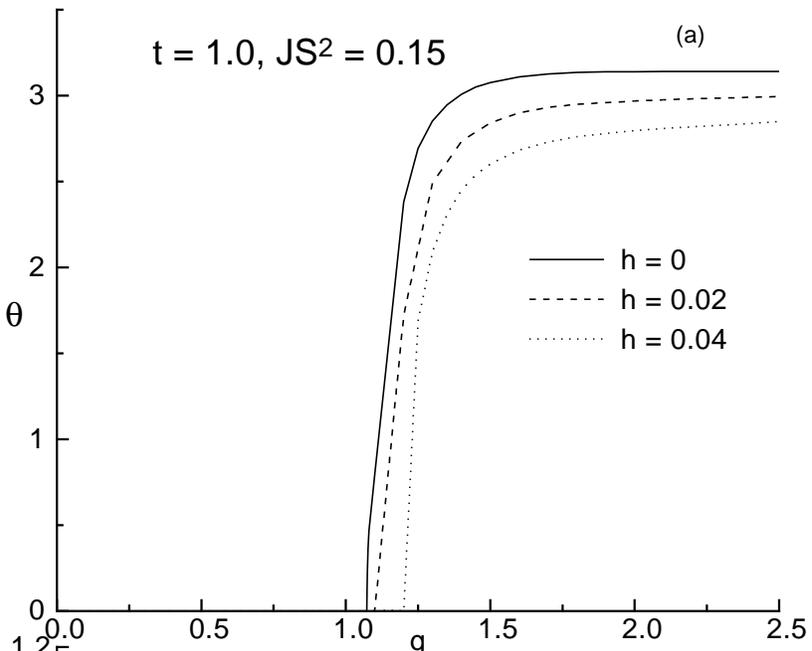
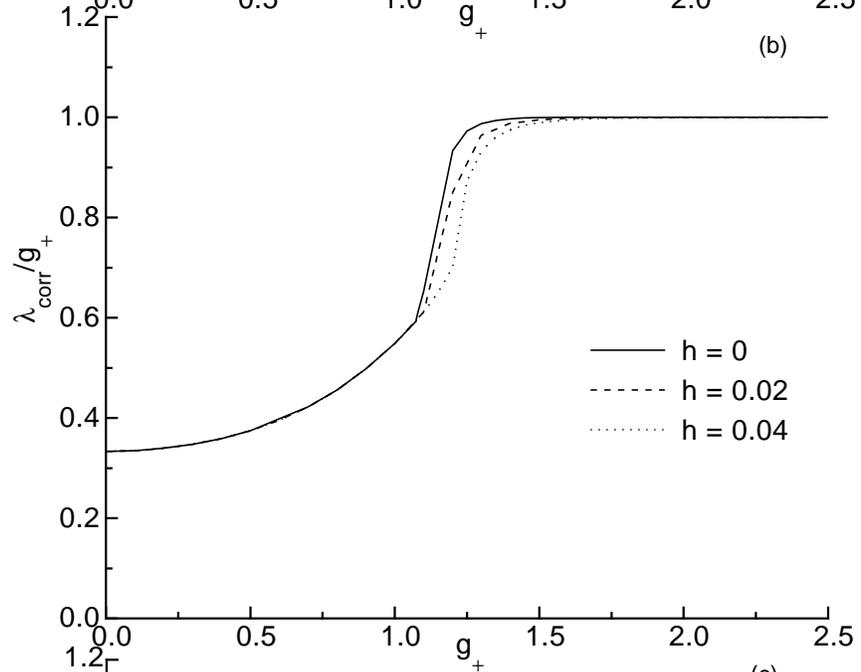
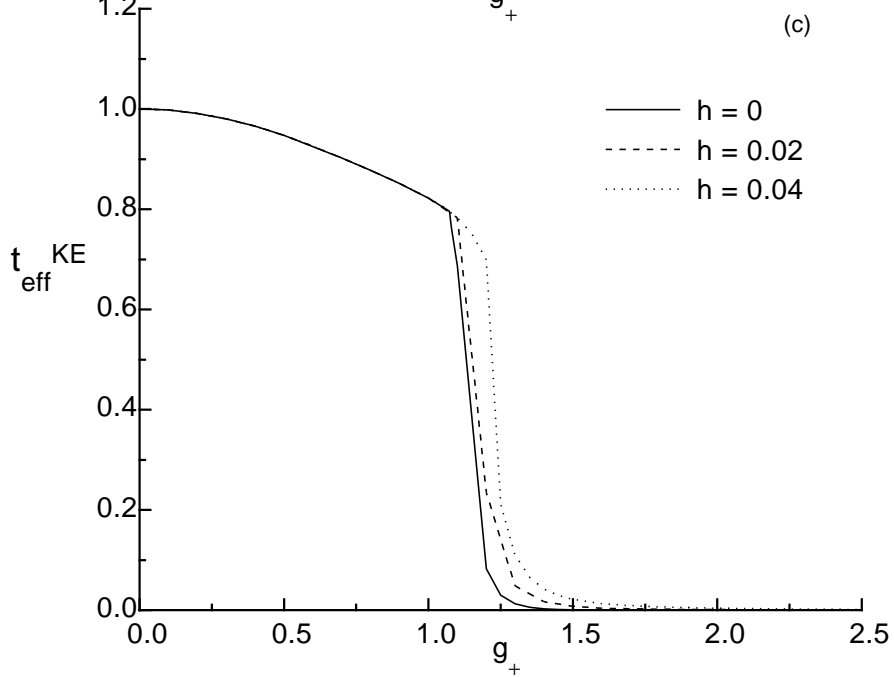
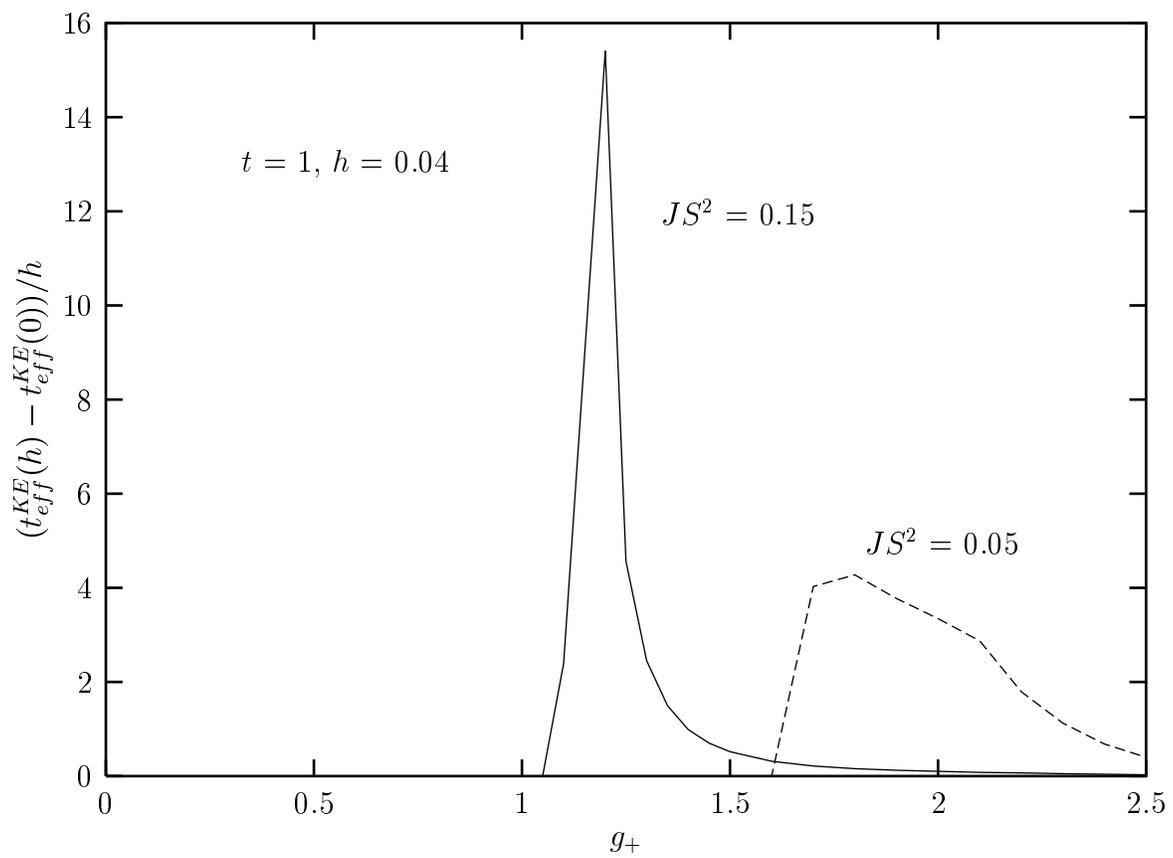
\end{document}